# MIMO Z Channel Interference Management

Ian Lim[1], Chedd Marley[2], and Jorge Kitazuru[3]

[1]National University of Singapore, Singapore
`ianlimsg@gmail.com`
[2]University of Sydney, Sydney, Australia
[3]University of Auckland, Auckland, New Zealand

*ABSTRACT*

*MIMO Z Channel is investigated in this paper. We focus on how to tackle the interference when different users try to send their codewords to their corresponding receivers while only one user will cause interference to the other. We assume there are two transmitters and two receivers each with two antennas. We propose a strategy to remove the interference while allowing different users transmit at the same time. Our strategy is low-complexity while the performance is good. Mathematical analysis is provided and simulations are given based on our system.*

## 1. INTRODUCTION

An (N,M)-MIMO wireless system can be generally defined as a MIMO system in which N signals are transmitted by N antennas at the same time using the same bandwidth and, thanks to effective processing at the receiver side based on the M received signals by M different antennas, is able to distinguish the different transmitted signals. The processing at the receiver is essentially efficient co-channel interference cancellation on the basis of the collected multiple information. This permits improving system performance whether the interest is to increase the single link data rate or increase the number of users in the whole system [1–16].

Z interference channel is a network consisting 2 senders and 2 receivers. There exists a one-to-one correspondence between senders and receivers. Each sender only wants to communicate with its corresponding receiver, and each receiver only cares about the information form its corresponding sender. However, a channel with strong signals may interfere other channels with weak signals while the channel with weak signals will not cause interference to the channel with strong signals. So Z interference channel has two principal links and one interference links. This scenario often occurs, when several sender-receiver pairs share a common media. The study of this kind of channel was studied in many literatures [17, 20–29]. However, this channel has not been solved in general case even in the general Gaussian case.

In this paper, we focus on MIMO Z channels. Since each user transmits at the same time, how to deal with the co-channel interference is an interesting question. When channel knowledge is known at the transmitter, schemes to cancel the co-channel interference are proposed in [18, 19, 30–34]. In this paper, we propose and analyze a scheme when channel knowledge is not known at the transmitter, a scenario which is more practical. The article is organized as follows. In the next section the system model is introduced. Detailed interference cancellation procedures are provided and performance analysis is given. Then simulation results are presented. Concluding remarks are given in the final section.

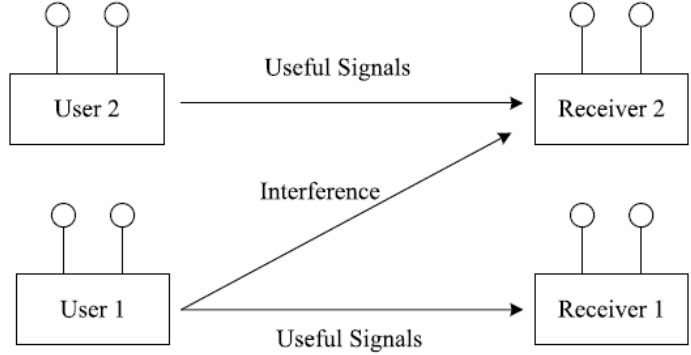

Figure 1: Channel Model

## 2. INTERFERENCE CANCELLATION AND PERFORMANCE ANALYSIS

Assume there are 2 transmitters each with 2 transmit antennas and 2 receivers each equipped with 2 receive antennas. Each transmitter sends codewords to different receivers. But only one user will cause interference to the other. So this is a MIMO Z channel. Let $c_{t,n}(j)$ denote the transmitted symbol from the $n$-th antenna of user $j$ at transmission interval $t$ and $r_{t,m}$ be the received word at the receive antenna $m$ at the first receiver. Let $\alpha_{i,k}(j)$ denote the channel elements from antenna $i$ to antenna $k$ for user $j$. Let $\eta_{i,j}$ denote the noise at the receiver. In order to reduce the number of required receive antennas, we propose a scheme to cancel the interference with less number of receive antennas.

Consider two users each transmitting Alamouti code, i.e. Orthogonal Space-Time
Block Code (OSTBC)
$$\begin{pmatrix} a_1 & a_2 \\ -a_2^* & a_1^* \end{pmatrix}$$
to receivers each equipped with 2 receive antennas.

We first consider receiver 2. The received signal at the first receive antenna can be written in the following format:

$$\begin{pmatrix} r_{1,1} \\ r_{2,1} \end{pmatrix} = \begin{pmatrix} s_1(1) & s_2(1) \\ -s_2(1)^* & s_1(1)^* \end{pmatrix} \begin{pmatrix} \alpha_{1,1}(1) \\ \alpha_{2,1}(1) \end{pmatrix} + \begin{pmatrix} s_1(2) & s_2(2) \\ -s_2(2)^* & s_1(2)^* \end{pmatrix} \begin{pmatrix} \alpha_{1,1}(2) \\ \alpha_{2,1}(2) \end{pmatrix} + \begin{pmatrix} \eta_{1,1} \\ \eta_{2,1} \end{pmatrix} \qquad (1)$$

At the second receive antenna, we have

$$\begin{pmatrix} r_{1,2} \\ r_{2,2} \end{pmatrix} = \begin{pmatrix} s_1(1) & s_2(1) \\ -s_2(1)^* & s_1(1)^* \end{pmatrix} \begin{pmatrix} \alpha_{1,2}(1) \\ \alpha_{2,2}(1) \end{pmatrix} + \begin{pmatrix} s_1(2) & s_2(2) \\ -s_2(2)^* & s_1(2)^* \end{pmatrix} \begin{pmatrix} \alpha_{1,2}(2) \\ \alpha_{2,2}(2) \end{pmatrix} + \begin{pmatrix} \eta_{1,2} \\ \eta_{2,2} \end{pmatrix} \qquad (2)$$

At the receiver 1, there is no interference, the received signal at the first antenna is

$$\begin{pmatrix} r_{1,1} \\ r_{2,1} \end{pmatrix} = \begin{pmatrix} s_1(1) & s_2(1) \\ -s_2(1)^* & s_1(1)^* \end{pmatrix} \begin{pmatrix} \alpha_{1,1}(1) \\ \alpha_{2,1}(1) \end{pmatrix} + + \begin{pmatrix} \eta_{1,1} \\ \eta_{2,1} \end{pmatrix} \qquad (3)$$

At the second receive antenna, we have

$$\begin{pmatrix} r_{1,2} \\ r_{2,2} \end{pmatrix} = \begin{pmatrix} s_1(1) & s_2(1) \\ -s_2(1)^* & s_1(1)^* \end{pmatrix} \begin{pmatrix} \alpha_{1,2}(1) \\ \alpha_{2,2}(1) \end{pmatrix} + + \begin{pmatrix} \eta_{1,2} \\ \eta_{2,2} \end{pmatrix} \qquad (4)$$

The idea behind interference cancellation arises from separate decodability of each symbol; at each receive antenna we perform the decoding algorithm as if there is only one user. This user will be the one the effect of whom we want to cancel out. Then, we simply subtract the soft-decoded value of each symbol in one of the receive antennas from the rest and as a result remove the effect of that user. This procedure is presented in the following. At the first antenna of receiver 2, we have

$$\begin{pmatrix} r_{1,1} \\ r_{2,1}^* \end{pmatrix} = \begin{pmatrix} \alpha_{1,1}(1) & \alpha_{2,1}(1) \\ \alpha_{2,1}(1)^* & -\alpha_{1,1}(1)^* \end{pmatrix} \begin{pmatrix} s_1(1) \\ s_2(1) \end{pmatrix} + \begin{pmatrix} \alpha_{1,1}(2) & \alpha_{2,1}(2) \\ \alpha_{2,1}(2)^* & -\alpha_{1,1}(2)^* \end{pmatrix} \begin{pmatrix} s_1(2) \\ s_2(2) \end{pmatrix} + \begin{pmatrix} \eta_{1,1} \\ \eta_{2,1}^* \end{pmatrix} \qquad (5)$$

At the second antenna of receiver 2, we have

$$\begin{pmatrix} r_{1,2} \\ r_{2,2}^* \end{pmatrix} = \begin{pmatrix} \alpha_{1,2}(1) & \alpha_{2,2}(1) \\ \alpha_{2,2}(1)^* & -\alpha_{1,2}(1)^* \end{pmatrix} \begin{pmatrix} s_1(1) \\ s_2(1) \end{pmatrix} + \begin{pmatrix} \alpha_{1,2}(2) & \alpha_{2,2}(2) \\ \alpha_{2,2}(2)^* & -\alpha_{1,2}(2)^* \end{pmatrix} \begin{pmatrix} s_1(2) \\ s_2(2) \end{pmatrix} + \begin{pmatrix} \eta_{1,2} \\ \eta_{2,2}^* \end{pmatrix} \qquad (6)$$

At the first antenna of receiver 1, we have

$$\begin{pmatrix} r_{1,1} \\ r_{2,1}^* \end{pmatrix} = \begin{pmatrix} \alpha_{1,1}(1) & \alpha_{2,1}(1) \\ \alpha_{2,1}(1)^* & -\alpha_{1,1}(1)^* \end{pmatrix} \begin{pmatrix} s_1(1) \\ s_2(1) \end{pmatrix} + \begin{pmatrix} \eta_{1,1} \\ \eta_{2,1}^* \end{pmatrix} \qquad (7)$$

At the second antenna of receiver 1, we have

$$\begin{pmatrix} r_{1,2} \\ r_{2,2}^* \end{pmatrix} = \begin{pmatrix} \alpha_{1,2}(1) & \alpha_{2,2}(1) \\ \alpha_{2,2}(1)^* & -\alpha_{1,2}(1)^* \end{pmatrix} \begin{pmatrix} s_1(1) \\ s_2(1) \end{pmatrix} + \begin{pmatrix} \eta_{1,2} \\ \eta_{2,2}^* \end{pmatrix} \qquad (8)$$

In order to cancel the signals $s11$ and $s12$ from User 1 at receiver 2, we first multiply both sides of Equation (25) with

$$\begin{pmatrix} \alpha_{1,1}(1) & \alpha_{2,1}(1) \\ \alpha_{2,1}(1)^* & -\alpha_{1,1}(1)^* \end{pmatrix}^\dagger$$

and multiply both sides of Equation (26) with

$$\begin{pmatrix} \alpha_{1,2}(1) & \alpha_{2,2}(1) \\ \alpha_{2,2}(1)^* & -\alpha_{1,2}(1)^* \end{pmatrix}^\dagger$$

Then we have Equations (13) and (14) in the next page, where $\eta\_1,1, \eta\_2,1, \eta\_1,2, \eta\_2,2$ are given by

$$\begin{pmatrix} \eta'_{1,1} \\ \eta'_{2,1} \end{pmatrix} = \begin{pmatrix} \alpha^*_{1,1}(1) & \alpha_{2,1}(1) \\ \alpha^*_{2,1}(1) & -\alpha_{1,1}(1) \end{pmatrix} \begin{pmatrix} \eta_{1,1} \\ \eta_{2,1} \end{pmatrix} \quad (9)$$

$$\begin{pmatrix} \eta'_{1,2} \\ \eta'_{2,2} \end{pmatrix} = \begin{pmatrix} \alpha^*_{1,2}(1) & \alpha_{2,2}(1) \\ \alpha^*_{2,2}(1) & -\alpha_{1,2}(1) \end{pmatrix} \begin{pmatrix} \eta_{1,2} \\ \eta^*_{2,2} \end{pmatrix} \quad (10)$$

In order to eliminate the effect of user 1, we need to divide both sides of Equation (13) by

$$\frac{1}{(|\alpha_{1,1}(1)|^2 + |\alpha_{2,1}(1)|^2)}$$

and divide both sides of Equation (14) by

$$\frac{1}{(|\alpha_{1,2}(1)|^2 + |\alpha_{2,2}(1)|^2)}$$

$$\begin{pmatrix} \alpha^*_{1,1}(1) & \alpha_{2,1}(1) \\ \alpha^*_{2,1}(1) & -\alpha_{1,1}(1) \end{pmatrix} \begin{pmatrix} r_{1,1} \\ r^*_{2,1} \end{pmatrix} = (|\alpha_{1,1}(1)|^2 + |\alpha_{2,1}(1)|^2) \begin{pmatrix} s^1_1 \\ s^1_2 \end{pmatrix}$$
$$+ \begin{pmatrix} \alpha^*_{1,1}(1) & \alpha_{2,1}(1) \\ \alpha^*_{2,1}(1) & -\alpha_{1,1}(1) \end{pmatrix} \begin{pmatrix} \alpha_{1,1}(2) & \alpha_{2,1}(2) \\ \alpha^*_{2,1}(2) & -\alpha^*_{1,1}(2) \end{pmatrix} \begin{pmatrix} s_1(2) \\ s_2(2) \end{pmatrix} + \begin{pmatrix} \eta'_{1,1} \\ \eta'_{2,1} \end{pmatrix} \quad (13)$$

$$\begin{pmatrix} \alpha^*_{1,2}(1) & \alpha_{2,2}(1) \\ \alpha^*_{2,2}(1) & -\alpha_{1,2}(1) \end{pmatrix} \begin{pmatrix} r_{1,2} \\ r^*_{2,2} \end{pmatrix} = (|\alpha_{1,2}(1)|^2 + |\alpha_{2,2}(1)|^2) \begin{pmatrix} s^1_1 \\ s^1_2 \end{pmatrix}$$
$$+ \begin{pmatrix} \alpha^*_{1,2}(1) & \alpha_{2,2}(1) \\ \alpha^*_{2,2}(1) & -\alpha_{1,2}(1) \end{pmatrix} \begin{pmatrix} \alpha_{1,2}(2) & \alpha_{2,2}(2) \\ \alpha^*_{2,2}(2) & -\alpha^*_{1,2}(2) \end{pmatrix} \begin{pmatrix} s_1(2) \\ s_2(2) \end{pmatrix} + \begin{pmatrix} \eta'_{1,2} \\ \eta'_{2,2} \end{pmatrix} \quad (14)$$

$$\frac{1}{(|\alpha_{1,1}(1)|^2 + |\alpha_{2,1}(1)|^2)} \begin{pmatrix} \alpha^*_{1,1}(1) & \alpha_{2,1}(1) \\ \alpha^*_{2,1}(1) & -\alpha_{1,1}(1) \end{pmatrix} \begin{pmatrix} r_{1,1} \\ r^*_{2,1} \end{pmatrix} = \begin{pmatrix} s^1_1 \\ s^1_2 \end{pmatrix} + \frac{1}{(|\alpha_{1,1}(1)|^2 + |\alpha_{2,1}(1)|^2)} \begin{pmatrix} \eta'_{1,1} \\ \eta'_{2,1} \end{pmatrix}$$
$$+ \left( \frac{1}{(|\alpha_{1,1}(1)|^2 + |\alpha_{2,1}(1)|^2)} \begin{pmatrix} \alpha^*_{1,1}(1) & \alpha_{2,1}(1) \\ \alpha^*_{2,1}(1) & -\alpha_{1,1}(1) \end{pmatrix} \begin{pmatrix} \alpha_{1,1}(2) & \alpha_{2,1}(2) \\ \alpha^*_{2,1}(2) & -\alpha^*_{1,1}(2) \end{pmatrix} \begin{pmatrix} s_1(2) \\ s_2(2) \end{pmatrix} \right) \quad (15)$$

$$\frac{1}{(|\alpha_{1,2}(1)|^2 + |\alpha_{2,2}(1)|^2)} \begin{pmatrix} \alpha^*_{1,2}(1) & \alpha_{2,2}(1) \\ \alpha^*_{2,2}(1) & -\alpha_{1,2}(1) \end{pmatrix} \begin{pmatrix} r_{1,2} \\ r^*_{2,2} \end{pmatrix} = \begin{pmatrix} s^1_1 \\ s^1_2 \end{pmatrix} + \frac{1}{(|\alpha_{1,2}(1)|^2 + |\alpha_{2,2}(1)|^2)} \begin{pmatrix} \eta'_{1,2} \\ \eta'_{2,2} \end{pmatrix}$$
$$+ \frac{1}{(|\alpha_{1,2}(1)|^2 + |\alpha_{2,2}(1)|^2)} \begin{pmatrix} \alpha^*_{1,2}(1) & \alpha_{2,2}(1) \\ \alpha^*_{2,2}(1) & -\alpha_{1,2}(1) \end{pmatrix} \begin{pmatrix} \alpha_{1,2}(2) & \alpha_{2,2}(2) \\ \alpha^*_{2,2}(2) & -\alpha^*_{1,2}(2) \end{pmatrix} \begin{pmatrix} s_1(2) \\ s_2(2) \end{pmatrix} \quad (16)$$

Equations (13) and (14) become Equations (15) and (16). Then we can subtract both sides of

Equation (15) from Equation (16). The resulting terms are shown by

$$\hat{y} = \hat{H}\begin{pmatrix}s_1(2)\\s_2(2)\end{pmatrix} + \begin{pmatrix}\eta''_{1,2}\\\eta''_{2,2}\end{pmatrix} \qquad (11)$$

where $\hat{y}$ and $\hat{H}$ are given by Equations (17) and (18). $\eta\_\_1,2, \eta\_\_2,2$ are given by

$$\begin{pmatrix}\eta''_{1,2}\\\eta''_{2,2}\end{pmatrix} = \frac{1}{(|\alpha_{1,2}(1)|^2 + |\alpha_{2,2}(1)|^2)}\begin{pmatrix}\eta'_{1,2}\\\eta'_{2,2}\end{pmatrix}$$

$$- \frac{1}{(|\alpha_{1,1}(1)|^2 + |\alpha_{2,1}(1)|^2)}\begin{pmatrix}\eta'_{1,1}\\\eta'_{2,1}\end{pmatrix} \qquad (12)$$

The distribution of $\eta\_\_1,2, \eta\_\_2,2$ are Gaussian white noise. In Equation (13), $\hat{H}$ can be written as the following structure:

$$\hat{H} = \begin{pmatrix} a & b \\ b^* & -a^* \end{pmatrix} \qquad (21)$$

where *a* and *b* are given by Equations (19) and (20). In order to decode the $s_1^2$, we can multiply both sides of the Equation (11) with

$$\begin{pmatrix}a\\b^*\end{pmatrix}^\dagger$$

$$\begin{pmatrix}a\\b^*\end{pmatrix}^\dagger \hat{y} = \begin{pmatrix}|a|^2 + |b|^2 & 0\end{pmatrix}\begin{pmatrix}s_1(2)\\s_2(2)\end{pmatrix}$$

$$+ \begin{pmatrix}a\\b^*\end{pmatrix}^\dagger \begin{pmatrix}\eta''_{1,2}\\\eta''_{2,2}\end{pmatrix}$$

$$= (|a|^2 + |b|^2)s_1(2) + \begin{pmatrix}a\\b^*\end{pmatrix}^\dagger \begin{pmatrix}\eta''_{1,2}\\\eta''_{2,2}\end{pmatrix} \qquad (22)$$

$$\widehat{y} = \frac{1}{|\alpha_{1,2}(1)|^2 + |\alpha_{2,2}(1)|^2} \begin{pmatrix} \alpha_{1,2}^*(1) & \alpha_{2,2}(1) \\ \alpha_{2,2}^*(1) & -\alpha_{1,2}(1) \end{pmatrix} \begin{pmatrix} r_{1,2} \\ r_{2,2}^* \end{pmatrix}$$
$$- \frac{1}{|\alpha_{1,1}(1)|^2 + |\alpha_{2,1}(1)|^2} \begin{pmatrix} \alpha_{1,1}^*(1) & \alpha_{2,1}(1) \\ \alpha_{2,1}^*(1) & -\alpha_{1,1}(1) \end{pmatrix} \begin{pmatrix} r_{1,1} \\ r_{2,1}^* \end{pmatrix} \quad (17)$$

$$\widehat{H} = \left[ \frac{1}{|\alpha_{1,2}(1)|^2 + |\alpha_{2,2}(1)|^2} \begin{pmatrix} \alpha_{1,2}^*(1) & \alpha_{2,2}(1) \\ \alpha_{2,2}^*(1) & -\alpha_{1,2}(1) \end{pmatrix} \begin{pmatrix} \alpha_{1,2}(2) & \alpha_{2,2}(2) \\ \alpha_{2,2}^*(2) & -\alpha_{1,2}^*(2) \end{pmatrix} \right.$$
$$\left. - \frac{1}{|\alpha_{1,1}(1)|^2 + |\alpha_{2,1}(1)|^2} \begin{pmatrix} \alpha_{1,1}^*(1) & \alpha_{2,1}(1) \\ \alpha_{2,1}^*(1) & -\alpha_{1,1}(1) \end{pmatrix} \begin{pmatrix} \alpha_{1,1}(2) & \alpha_{2,1}(2) \\ \alpha_{2,1}^*(2) & -\alpha_{1,1}^*(2) \end{pmatrix} \right] (18)$$

$$a = \frac{1}{|\alpha_{1,2}(1)|^2 + |\alpha_{2,2}(1)|^2} [\alpha_{1,2}^*(1)\alpha_{1,2}(2) + \alpha_{2,2}(1)\alpha_{2,2}^*(2)]$$
$$- \frac{1}{|\alpha_{1,1}(1)|^2 + |\alpha_{2,1}(1)|^2} [\alpha_{1,1}^*(1)\alpha_{1,1}(2) + \alpha_{2,1}(1)\alpha_{2,1}^*(2)] \quad (19)$$

$$b = \frac{1}{|\alpha_{1,2}(1)|^2 + |\alpha_{2,2}(1)|^2} [\alpha_{1,2}^*(1)\alpha_{2,2}(2) - \alpha_{2,2}(1)\alpha_{1,2}^*(2)]$$
$$- \frac{1}{|\alpha_{1,1}(1)|^2 + |\alpha_{2,1}(1)|^2} [\alpha_{1,1}^*(1)\alpha_{2,1}(2) - \alpha_{2,1}(1)\alpha_{1,1}^*(2)] \quad (20)$$

In order to keep the Gaussian white noise, we need

$$\frac{1}{\sqrt{|a|^2 + |b|^2}} \begin{pmatrix} a \\ b^* \end{pmatrix}^\dagger \widehat{y} = \sqrt{|a|^2 + |b|^2} s_1(2)$$
$$+ \frac{1}{\sqrt{|a|^2 + |b|^2}} \begin{pmatrix} a \\ b^* \end{pmatrix}^\dagger \begin{pmatrix} \eta_{1,2}'' \\ \eta_{2,2}'' \end{pmatrix} \quad (23)$$

Maximum likelihood decoding can be used to decode $s_1^2$

$$\widehat{s}_1^2 = \arg\min_{s_1^2} \left| \frac{1}{\sqrt{|a|^2 + |b|^2}} \begin{pmatrix} a \\ b^* \end{pmatrix}^\dagger \widehat{y} - \sqrt{|a|^2 + |b|^2} s_1(2) \right|_F^2 \quad (24)$$

So the decoding is symbol-by-symbol. We know at the first antenna of receiver 1, we have

$$\begin{pmatrix} r_{1,1} \\ r_{2,1}^* \end{pmatrix} = \begin{pmatrix} \alpha_{1,1}(1) & \alpha_{2,1}(1) \\ \alpha_{2,1}(1)^* & -\alpha_{1,1}(1)^* \end{pmatrix} \begin{pmatrix} s_1(1) \\ s_2(1) \end{pmatrix} + \begin{pmatrix} \eta_{1,1} \\ \eta_{2,1}^* \end{pmatrix} \quad (25)$$

At the second antenna of receiver 1, we have

By multiplying (25) with

$$\begin{pmatrix} \alpha_{1,1}(1) & \alpha_{2,1}(1) \\ \alpha_{2,1}(1)^* & -\alpha_{1,1}(1)^* \end{pmatrix}^\dagger \qquad (27)$$

and multiplying (26) with

$$\begin{pmatrix} \alpha_{1,2}(1) & \alpha_{2,2}(1) \\ \alpha_{2,2}(1)^* & -\alpha_{1,2}(1)^* \end{pmatrix}^\dagger \qquad (28)$$

we can decode signals from user 1 to receiver 1. Now we analyze the diversity. From Equation (22), we know that the diversity is determined by factor $\sqrt{|a|^2 + |\tilde{b}|^2}$. diversity is defined as

$$d = -\lim_{\rho \to \infty} \frac{\log P_e}{\log \rho} \qquad (29)$$

where $\rho$ denotes the SNR and $P_e$ represents the probability of error. It is known that the error probability can be written as

$$\begin{aligned} & P(s_1(2) \to error|a,b) \\ &= Q\left(\sqrt{\frac{\rho|\sqrt{|a|^2+|b|^2}\mathbf{e}|_F^2}{4}}\right) \\ &\le \exp\left(-\frac{\rho(|a|^2+|b|^2)\mathbf{e}^\dagger \mathbf{e}}{4}\right) \\ &= \exp\left(-\frac{\rho(|a|^2+|b|^2)e^2}{4}\right) \end{aligned} \qquad (30)$$

where $e$ is the error. We need to analyze $a$ and $b$. Conditioned on $\alpha 1,2(1)$, $\alpha 2,2(1)$, $\alpha 1,1(1)$, $\alpha 2,1(1)$, then $a$ and $b$ are both Gaussian random variables. It is easy to verify that

$$E[a \cdot b | \alpha_{1,2}(1), \alpha_{2,2}(1), \alpha_{1,1}(1), \alpha_{2,1}(1)] = 0 \qquad (31)$$

So $a$ and $b$ are independent Gaussian random variables. We have

$$\begin{aligned} & P(s_1(2) \to error) \\ &= E[E[P(s_1(2) \to error|a,b)]| \\ & \qquad \alpha_{1,2}(1), \alpha_{2,2}(1), \alpha_{1,1}(1), \alpha_{2,1}(1)] \\ &\le E[E[\exp\left(-\frac{\rho(|a|^2+|b|^2)e^2}{4}\right)| \\ & \qquad \alpha_{1,2}(1), \alpha_{2,2}(1), \alpha_{1,1}(1), \alpha_{2,1}(1)]] \\ &= E[\frac{1}{\prod_{j=1}^2 [1+\frac{\rho e^2}{4})]}| \\ & \qquad \alpha_{1,2}(1), \alpha_{2,2}(1), \alpha_{1,1}(1), \alpha_{2,1}(1)] \\ &= \frac{1}{\prod_{j=1}^2 [1+\frac{\rho e^2}{4})]} \end{aligned} \qquad (32)$$

When $\rho$ is large, Equation (32) becomes

$$P(s_1(2) \to error) \leq \rho^{-2} \left(\frac{e^2}{4}\right)^{-2} \qquad (33)$$

By Equation (29), the diversity is 2. Now we analyze the diversity for $s2(2)$. We know that the diversity is determined by factor $\sqrt{|a|^2 + |b|^2}$. The error probability can be written as

$$\begin{aligned}
P(s_2(2) &\to error|a,b) \\
&= Q\left(\sqrt{\frac{\rho|\sqrt{|a|^2+|b|^2}\mathbf{e}|_F^2}{4}}\right) \\
&\leq \exp\left(-\frac{\rho(|a|^2+|b|^2)\mathbf{e}^\dagger\mathbf{e}}{4}\right) \\
&= \exp\left(-\frac{\rho(|a|^2+|b|^2)e^2}{4}\right)
\end{aligned} \qquad (34)$$

where $e$ is the error. We need to analyze $a$ and $b$. Conditioned on $\alpha_{1,2}(2)$, $\alpha_{2,2}(2)$, $\alpha_{1,1}(2)$, $\alpha_{2,1}(2)$, then $a$ and $b$ are both Gaussian random variables. It is easy to verify that

$$E[a \cdot b | \alpha_{1,2}(2), \alpha_{2,2}(2), \alpha_{1,1}(2), \alpha_{2,1}(2)] = 0 \qquad (35)$$

So $a$ and $b$ are independent Gaussian random variables. We have

$$\begin{aligned}
P(s_2(2) &\to error) \\
&= E[E[P(s_2(2) \to error|a,b)]| \\
&\qquad \alpha_{1,2}(2), \alpha_{2,2}(2), \alpha_{1,1}(2), \alpha_{2,1}(2)] \\
&\leq E[E[\exp\left(-\frac{\rho(|a|^2+|b|^2)e^2}{4}\right)| \\
&\qquad \alpha_{1,2}(2), \alpha_{2,2}(2), \alpha_{1,1}(2), \alpha_{2,1}(2)]] \\
&= E[\frac{1}{\prod_{j=1}^2 [1+\frac{\rho e^2}{4})]}| \\
&\qquad \alpha_{1,2}(2), \alpha_{2,2}(2), \alpha_{1,1}(2), \alpha_{2,1}(2)] \\
&= \frac{1}{\prod_{j=1}^2 [1+\frac{\rho e^2}{4})]}
\end{aligned} \qquad (36)$$

When $\rho$ is large, Equation (36) becomes

$$P(s_2(2) \to error) \leq \rho^{-2} \left(\frac{e^2}{4}\right)^{-2} \qquad (37)$$

By Equation (29), the diversity for $s_2^2$ is 2. Similarly, for receiver 1, we have

$$\begin{aligned}
&P(s_2(1) \to error) \\
&= E[E[P(s_2(1) \to error|a,b)]| \\
&\quad \alpha_{1,2}(1), \alpha_{2,2}(1), \alpha_{1,1}(1), \alpha_{2,1}(1)] \\
&\leq E[E[\exp\left(-\frac{\rho(|a|^2+|b|^2)e^2}{4}\right)| \\
&\quad \alpha_{1,2}(1), \alpha_{2,2}(1), \alpha_{1,1}(1), \alpha_{2,1}(1)]] \\
&= E[\frac{1}{\prod_{j=1}^{2}[1+\frac{\rho e^2}{4})]}| \\
&\quad \alpha_{1,2}(1), \alpha_{2,2}(1), \alpha_{1,1}(1), \alpha_{2,1}(1)] \\
&= \frac{1}{\prod_{j=1}^{2}[1+\frac{\rho e^2}{4})]}
\end{aligned} \qquad (38)$$

When $\rho$ is large, Equation (36) becomes

$$P(s_2(1) \to error) \leq \rho^{-2}\left(\frac{e^2}{4}\right)^{-2} \qquad (39)$$

By Equation (29), the diversity for *s*2(1) is 2.

In summary, for receiver 2, the interference cancellation based on Alamouti codes can achieve cancel the interference successfully and the decoding complexity is symbolby- symbol which is the lowest and the diversity is 2. For receiver 1, the interference cancellation based on Alamouti codes can achieve cancel the interference successfully and the decoding complexity is symbol-by-symbol which is the lowest and the diversity is 2.

## 3. SIMULATIONS

In order to evaluate the proposed scheme, we use a system with two users with two antennas and two receivers each with two receive antennas. This is a typical MIMO Z channel. The two users are sending signals to the receivers simultaneously. We assume alamouti codes are transmitted. So there will be co-channel interference. If the proposed interference cancellation is used, the performance is provided in Figures 2 while QPSK is used in Figure 2. In figure 2, we compare the interference cancellation scheme with a TDMA scheme with beamforming scheme. That is, during each time slot, one user transmits while the other keeps silent. In order to make the rate the same for the two schemes, in Figure 2, 16-QAM is used. It is obvious that the proposed scheme has better performance which confirms the effectiveness of the interference cancellation scheme.

## 4. CONCLUSIONS

In this paper, we discuss the MIMO Z channel. We first give detailed description on MIMO Z channel. Later we show that how to tackle interference in such a system is important. Aiming to remove the interference, a strategy for MIMO Z channel is proposed and analyzed. We assume there are two transmitters and two receivers each with two antennas. The complexity of the strategy is low while the performance is good. Simulations confirm the theoretical analysis.

We therefore focus on the first term. Using the rule (10.17) for matrix differentiation [29], we

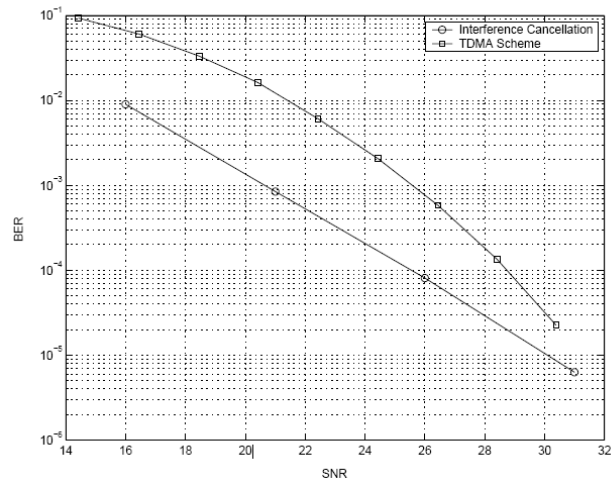

Figure 2: QPSK constellation with interference cancellation

## REFERENCE

[1] H. Jiao and F. Y. Li, "A TDMA-Based MAC Protocol Supporting Cooperative Communications in Wireless Mesh Networks," *International Journal of Computer Networks & Communications (IJCNC)*, 2011

[2] D. S Saini and N. Sharma, "Performance Improvement in OVSF Based CDMA Networks Using Flexible Assignment of Data Calls," *International Journal of Computer Networks & Communications (IJCNC)*, 2011.

[3] Khaled Day, Abderezak Touzene, Bassel Arafeh and Nasser Alzeidi, "Parallel Routing in Mobile Ad-Hoc Networks," *International Journal of Computer Networks & Communications (IJCNC)*, 2011

[4] L. Chaari and L. Kamoun, "Performance Analysis of IEEE 802.15.4/Zigbee Standard Under Real Time Constraints," *International Journal of Computer Networks & Communications (IJCNC)*, 2011.

[5] M. Chiani, M. Z. Win, H. Shin, "MIMO Networks: The Effects of Interference," *IEEE Transactions on Information Theory*, vol. 56, no. 1, pp. 336-349, 2010

[6] F. Li, "Optimization of input covariance matrix for multi-antenna correlated channels," *International Journal of Computer Networks & Communications (IJCNC)*, 2011.

[7] W. Y. Ge, J. S. Zhang and G. L. Xue, "MIMO-Pipe Modeling and Scheduling for Efficient Interference Management in Multihop MIMO Networks," *IEEE Transactions on Vehicular Technology*, vol. 59, no. 8, pp. 3966 3978, 2010.

[8] F. Li and H. Jafarkhani, "Space-Time Processing for X Channels Using Precoders," *IEEE Transactions on Signal Processing*.

[9] C. F. Ball, R. Mullner, J. Lienhart, H. Winkler, "Performance analysis of Closed and Open loop MIMO in LTE," *European Wireless Conference*, pp. 260-265, 2009

[10] H. Futaki, T. Ohtsuki, "Low-density parity-check (LDPC) coded MIMO systems with iterative turbo decoding," *Vehicular Technology Conference*, pp 342-346, 2003.

[11] A. Goldsmith, S. Jafar, N. Jindal, S. Vishwanath, " Capacity limits of MIMO channels," *IEEE Journal on Selected Areas in Communications*, vol. 21, no. 5, pp. 684-702, 2003.

[12] F. Li, Q. T. Zhang, and S. H. Song, "Efficient optimization of input covariance matrix for MISO in correlated Rayleigh fading," in *Proceedings of IEEE Wireless Communications and Networking Conference*, March 2007.

[13] C. Y. Chi and C. H. Chen, "Cumulant-based inverse filter criteria for MIMO blind deconvolution: properties, algorithms, and application to DS/CDMA systems in multipath," *IEEE Transactions on Signal Processing*, vol 49, no. pp. 1282-1299, 2001.

[14] F. Li, *Multi-Antenna Multi-User Interference Cancellation and Detection Using Precoders*, PhD thesis, UC Irvine, 2012

[15] K. Kusume, G. Dietl, T. Abe, H. Taoka, S. Nagata, "System Level Performance of Downlink MU-MIMO Transmission for 3GPP LTE-Advanced," *Vehicular Technology Conference*, 2010

[16] F. Li and H. Jafarkhani, "Interference Cancellation and Detection Using Precoders," *IEEE International Conference on Communications (ICC 2009)*, June 2009.

[17] F. Fernandes, S. Vishwanath, "On the capacity of degraded MIMO Z interference channels with degraded message sets," *Conference Record of the Forty Fourth Asilomar Conference on Signals, Systems and Computers (ASILOMAR)*, 2010.

[18] E. M. Mohamed, D. Kinoshita, K. Mitsunaga, Y. Higa, H. Furukawa, "MIMO based wireless backhaul," *Ultra Modern Telecommunications and Control Systems and Workshops (ICUMT)*, 2010.

[19] F. Li and H. Jafarkhani, "Multiple-antenna interference cancellation and detection for two users using precoders," *IEEE Journal of Selected Topics in Signal Processing*, December 2009.

[20] L. Ke; Z. Wang, "On the Degrees of Freedom Regions of Two-User MIMO Z and Full Interference Channels with Reconfigurable Antennas," IEEE Global Telecommunications Conference, 2010

[21] S. Karmakar, M. K. Varanasi, "The diversity-multiplexing tradeoff of the MIMO Z interference channel," *IEEE International Symposium on Information Theory Proceedings (ISIT)*, 2010

[22] F. Li and H. Jafarkhani, "Interference Cancellation and Detection for More than Two Users," *IEEE Transactions on Communications*, March 2011.

[23] F. Li and H. Jafarkhani, "Multiple-antenna interference cancellation and detection for two users using quantized feedback," *IEEE Transactions on Wireless Communication*, vol. 10, no. 1, pp. 154-163, Jan 2011.

[24] Y. Peng, D. Rajan, "Capacity Bounds for a Cognitive MIMO Gaussian Z-Interference Channel," *IEEE Transactions on Vehicular Technology*, vol. 59, no. 4, pp. 1865-1876, 2010.

[25] X. Shang, B. Chen, G. Kramer H. V. Poor, "MIMO Z-interference channels: Capacity under strong and noisy interference," *Conference Record of the Forty-Third Asilomar Conference on Signals, Systems and Computers*, 2009.

[26] F. Li and H. Jafarkhani, "Interference cancellation and detection for multiple access channels with four users," in *Proceedings of IEEE International Conference on Communications (ICC 2010)*, June 2010.

[27] F. Li and H. Jafarkhani, "Using quantized feedback to cancel interference in multiple access channels," in *Proceedings of IEEE Global Telecommunications Conference (Globecom 2010)*, December, 2010.

[28] H. T. Do, T. J. Oechtering, M. Skoglund, "Coding for the Z channel with a digital relay link," *2010 IEEE Information Theory Workshop (ITW)*, 2010.

[29] L. G. Tallini, S. Al-Bassam, B. Bose, "Feedback Codes Achieving the Capacity of the Z-Channel," *IEEE Transactions on Information Theory*, vol. 54, no. 3, pp. 1357- 1362, 2008.

[30] D. J. Ryan, I. B. Collings, I. V. L. Clarkson, and R. W. Heath, Jr., "Performance of Vector Perturbation Multiuser MIMO Systems with Limited Feedback," *IEEE Trans. on Communications*, vol. 57, no. 9, pp. 2633-2644, Sept. 2009.

[31] F. Li and Q. T. Zhang, "Transmission strategy for MIMO correlated rayleigh fading channels with mutual coupling," in *Proceedings of IEEE International Conference on Communications (ICC 2007)*, June, 2007.

[32] F. Li and H. Jafarkhani, "Resource allocation algorithms with reduced complexity in MIMO multi-hop fading channels," in *Proceedings of IEEE Wireless Communications and Networking Conference*, 2009.

[33] K. Huang, R. W. Heath, Jr., and J. G. Andrews, "Uplink SDMA with Limited Feedback: Throughput Scaling," *EURASIP Journal on Advances in Signal Processing, special issue on Limited Feedback*, vol. 2008, Article ID 479357, 2008.

[34] K. Huang, R. W. Heath, Jr., and J. G. Andrews, "Space Division Multiple Access with a Sum Feedback Rate Constraint," *IEEE Trans. on Signal Processing*, vol. 55, no. 7, pp. 3879-3891, July 2007.